\documentclass[10pt,onecolumn,preprintnumbers,amsmath,amssymb,nofootinbib,superscriptaddress,aps,prd,numberedappendix]{openjournal}

\usepackage{newtxtext,newtxmath}
\usepackage{lipsum}
\usepackage[T1]{fontenc}
\usepackage{ae,aecompl}

\usepackage{graphicx}	
\usepackage{amsmath}	
\usepackage{amssymb}	

\begin{document}

\title{Propagating Residual Biases in Masked Cosmic Shear Power Spectra}

\author{T. D. Kitching$^{1,\dagger}$ 
, A. C. Deshpande$^{1}$, P. L. Taylor$^{2}$}
\email{$^{\dagger}$t.kitching@ucl.ac.uk, © 2020. All rights reserved.}
\affiliation{
$^{1}$Mullard Space Science Laboratory, University College London, Holmbury St Mary, Dorking, Surrey RH5 6NT, UK\\
$^{2}$Jet Propulsion Laboratory, California Institute of Technology, 4800 Oak Grove Drive, Pasadena, CA 91109, USA}

\begin{abstract}
In this paper we derive a full expression for the propagation of weak lensing shape measurement biases into cosmic shear power spectra including the effect of missing data. We show using simulations that terms higher than first order in bias parameters can be ignored and the impact of biases can be captured by terms dependent only on the mean of the multiplicative bias field. We identify that the B-mode power contains information on the multiplicative bias. We find that without priors on the residual multiplicative bias $\delta m$ and stochastic ellipticity variance $\sigma_e$ that constraints on the amplitude of the cosmic shear power spectrum are completely degenerate, and that when applying priors the constrained amplitude $A$ is slightly biased low via a classic marginalisation paradox. Using all-sky Gaussian random field simulations we find that the combination of $(1+2\delta m)A$ is unbiased for a joint EE and BB power spectrum likelihood if the error and mean (precision and accuracy) of the stochastic ellipticity variance is known to better than $\sigma(\sigma_e)\leq 0.05$ and $\Delta\sigma_e\leq 0.01$, or the multiplicative bias is known to better than $\sigma(m)\leq 0.07$ and $\Delta m\leq 0.01$. 
\end{abstract}

\maketitle

\section{Introduction}
\label{S:Intro}
Measurements of the weak lensing effect can be subject to biases caused by inaccuracies in the algorithmic methods used to determine a galaxy's shape \citep{step1,step2,great08,great10,great3}, measure the point spread function of a system \citep{2017MNRAS.468.3295H,2019A&A...624A..92K}, determine detector effects \citep{2014JInst...9C3048A}, or detect galaxies \citep{cccp}. 

The treatment of such biases in cosmic shear power spectra is a topic that has been dealt with in several papers, for example \cite{AR,massey,cropper,K19}. However a full propagation of biases into measured (observed) power spectra in the presence of survey masks (where a portion of the sky is unobserved) has not been done. In this paper we build on the work of \cite{K19} and include the impact of survey masks. 

In Section \ref{S:Method} we present the formalism for propagation of biases in the presence of masks, and identify power spectrum combinations that are dependent to varying degrees on real and imaginary bias terms; in Section \ref{Simple Simulations} we test our formalism on simulations; we discuss conclusions in Section \ref{Conclusions}.
\\
\section{Method}
\label{S:Method}
In the following we expand upon the derivations given in \cite{K19}. We can relate a measured shear in real (angular) space to the true shear -- that would have been measured in the absence of systematic effects or a mask -- via multiplicative and additive fields that describe respective biases that may be introduced 
\begin{eqnarray}
    \label{gamma}
    \widetilde\gamma(\mathbf{\Omega})&=&W(\mathbf{\Omega})\widehat\gamma(\mathbf{\Omega})\nonumber\\
    \widetilde\gamma(\mathbf{\Omega})&=&W(\mathbf{\Omega})\{[1+m_0(\mathbf{\Omega})]\gamma(\mathbf{\Omega})+m_4(\mathbf{\Omega})\gamma^*(\mathbf{\Omega})+c(\mathbf{\Omega})\},
\end{eqnarray}
where all quantities are a function of angular coordinates $\mathbf{\Omega}=(\theta,\phi)$, with $\theta$ and $\phi$ being latitude and longitude (or R.A. and dec). $W(\mathbf{\Omega})$ is a spin-$0$ mask where $W(\mathbf{\Omega})=1$ where data exists and $W(\mathbf{\Omega})=0$ where there is no data (we note that an optimal weight could in principle be computed). $\gamma(\mathbf{\Omega})$ is the true spin-2 shear, $\widetilde\gamma(\mathbf{\Omega})$ is the measured spin-2 shear, and $\widehat\gamma(\mathbf{\Omega})$ is the measured spin-2 shear in the absence of a mask. $m_0(\mathbf{\Omega})=m_0^R(\mathbf{\Omega})+{\rm i}m_0^I(\mathbf{\Omega})$ is a position-dependent multiplicative bias term that includes a possible systematic rotation \citep[see][Appendix A]{K19}, $m_4(\mathbf{\Omega})=m_4^R(\mathbf{\Omega})+{\rm i}m_4^I(\mathbf{\Omega})$ is a possible spin-4 multiplicative bias term\footnote{We note that if one defines a multiplicative bias like $m_1(\mathbf{\Omega})\gamma_1(\mathbf{\Omega})+{\rm i}m_2(\mathbf{\Omega})\gamma_2(\mathbf{\Omega})$ then $m_1(\mathbf{\Omega})=m_2(\mathbf{\Omega})$ would imply that $m_4(\mathbf{\Omega})=0$, and $m_1(\mathbf{\Omega})=-m_2(\mathbf{\Omega})$ would imply that that $m_0(\mathbf{\Omega})=0$.}, and $c(\mathbf{\Omega})$ is a spin-2 position-dependent additive bias. $^*$ is a complex conjugate. 

The spherical harmonic coefficients for the E-mode (curl-free) and B-mode (divergence-free) parts of the shear field can be determined via 
\begin{eqnarray}
    \label{gtoe}
    \gamma^E_{\ell m}&=&\frac{1}{2}\int {\rm d}\mathbf{\Omega}\, [
    \gamma(\mathbf{\Omega})\,{}_2Y^*_{\ell m}(\mathbf{\Omega})+
    \gamma^*(\mathbf{\Omega})\,{}_{-2}Y^*_{\ell m}(\mathbf{\Omega})]\nonumber\\
    \gamma^B_{\ell m}&=&\frac{-{\rm i}}{2}\int {\rm d}\mathbf{\Omega}\, [
    \gamma(\mathbf{\Omega})\,{}_2Y^*_{\ell m}(\mathbf{\Omega})-
    \gamma^*(\mathbf{\Omega})\,{}_{-2}Y^*_{\ell m}(\mathbf{\Omega})], 
\end{eqnarray}
where ${}_sY_{\ell m}(\mathbf{\Omega})$ are spin-weighted spherical harmonics (with spin $s=2$ or $-2$), $\ell$ and $m$ are angular wavenumbers; note that we use $m$ as one of the spherical harmonic wavenumbers and $m(\mathbf{\Omega})$ as multiplicative biases to follow convention but these should not be confused. 

As shown in \cite{K19}, and including the spin-4 terms, in the absence of any mask (i.e. $W(\mathbf{\Omega})=1$ $\forall\;\mathbf{\Omega}$) the true and measured shear field's spherical harmonic coefficients can be written like 
\begin{eqnarray}
    \label{shtransform}
    \widehat\gamma^E_{\ell m}&=&
    \gamma^E_{\ell m}+\sum_{\ell'm'}
    [\gamma^E_{\ell'm'}({}^{m_0}W^+_{\ell\ell'm m'}+{}^{m_4}W^+_{\ell\ell'm m'})+
     \gamma^B_{\ell'm'}({}^{m_0}W^-_{\ell\ell'm m'}-{}^{m_4}W^-_{\ell\ell'm m'})]+
     c^E_{\ell m}\nonumber\\
    \widehat\gamma^B_{\ell m}&=&
    \gamma^B_{\ell m}+\sum_{\ell'm'}
    [\gamma^B_{\ell'm'}({}^{m_0}W^+_{\ell\ell'm m'}-{}^{m_4}W^+_{\ell\ell'm m'})-
     \gamma^E_{\ell'm'}({}^{m_0}W^-_{\ell\ell'm m'}+{}^{m_4}W^-_{\ell\ell'm m'})]+
     c^B_{\ell m}.
\end{eqnarray}
where we expand the spin-2 quantities like $f(\mathbf{\Omega})=\sum_{\ell m}f_{\ell m} \,{}_{2}Y_{\ell m}(\mathbf{\Omega})$. Throughout we use superscript and subscript labels, but we note that the position of the labels relative to the main symbol is not significant i.e. they are just labels. We note that the multiplicative weight factors for the $E$-mode parts only depend on the sum of the multiplicative bias terms.

The weight functions are given by 
\begin{eqnarray}
\label{W1}
    {}^mW^+_{\ell\ell' m m'}&=&\frac{1}{2}
    (_{2}W^{m^R,mm'}_{\ell\ell'}+_{-2}W^{m^R,mm'}_{\ell\ell'})+
    {\rm i}(_{2}W^{m^I,mm'}_{\ell\ell'}-_{-2}W^{m^I,mm'}_{\ell\ell'})]\nonumber\\
    {}^mW^-_{\ell\ell' m m'}&=&\frac{{\rm i}}{2}
    (_{2}W^{m^R,mm'}_{\ell\ell'}-_{-2}W^{m^R,mm'}_{\ell\ell'})+
    {\rm i}(_{2}W^{m^I,mm'}_{\ell\ell'}+_{-2}W^{m^I,mm'}_{\ell\ell'})],
\end{eqnarray}
for $m=m_0$ or $m_4$. 

In a similar way the impact of the mask can be related to the measurement in the absence of a mask using a standard pseudo-$C_{\ell}$ expression \citep{lct, zs, grain, bct} 
\begin{eqnarray}
    \label{shtransform2}
    \widetilde\gamma^E_{\ell m}&=&
    \sum_{\ell'm'}
    [\widehat\gamma^E_{\ell'm'}{}^wW^+_{\ell\ell'm m'}+
     \widehat\gamma^B_{\ell'm'}{}^wW^-_{\ell\ell'm m'}]\nonumber\\
    \widetilde\gamma^B_{\ell m}&=&
    \sum_{\ell'm'}
    [\widehat\gamma^B_{\ell'm'}{}^wW^+_{\ell\ell'm m'}-
     \widehat\gamma^E_{\ell'm'}{}^wW^-_{\ell\ell'm m'}].
\end{eqnarray}
The weight functions within the sums represent the mode-mixing caused by the mask and are given by  
\begin{eqnarray}
\label{W2}
    {}^wW^+_{\ell\ell' m m'}&=&\frac{1}{2}[
    (_{2}W^{W,mm'}_{\ell\ell'}+_{-2}W^{W,mm'}_{\ell\ell'})\nonumber\\
    {}^wW^-_{\ell\ell' m m'}&=&\frac{{\rm i}}{2}[
    (_{2}W^{W,mm'}_{\ell\ell'}-_{-2}W^{W,mm'}_{\ell\ell'}),
\end{eqnarray}
note that for the spin-0 mask the additional imaginary terms do not exist, but that all of these quantities are complex. These quantities in equations (\ref{W1}) and (\ref{W2}) are formed from combinations of integrals on the sphere over the mask or multiplicative bias field multiplied by spin-weighted spherical harmonic functions and are given by  
\begin{eqnarray}
\label{wigs}
    _{s}W^{f,mm'}_{\ell\ell'}=\int{\rm d}\mathbf{\Omega}\, _{s}Y^*_{\ell'm'}(\mathbf{\Omega}) f(\mathbf{\Omega}) _{s}Y_{\ell m}(\mathbf{\Omega});
\end{eqnarray}
where the function on the sphere $f(\mathbf{\Omega})$ in the integrand is labelled in the superscript. 
\begin{figure*}
\centering
\includegraphics[width=0.32\columnwidth]{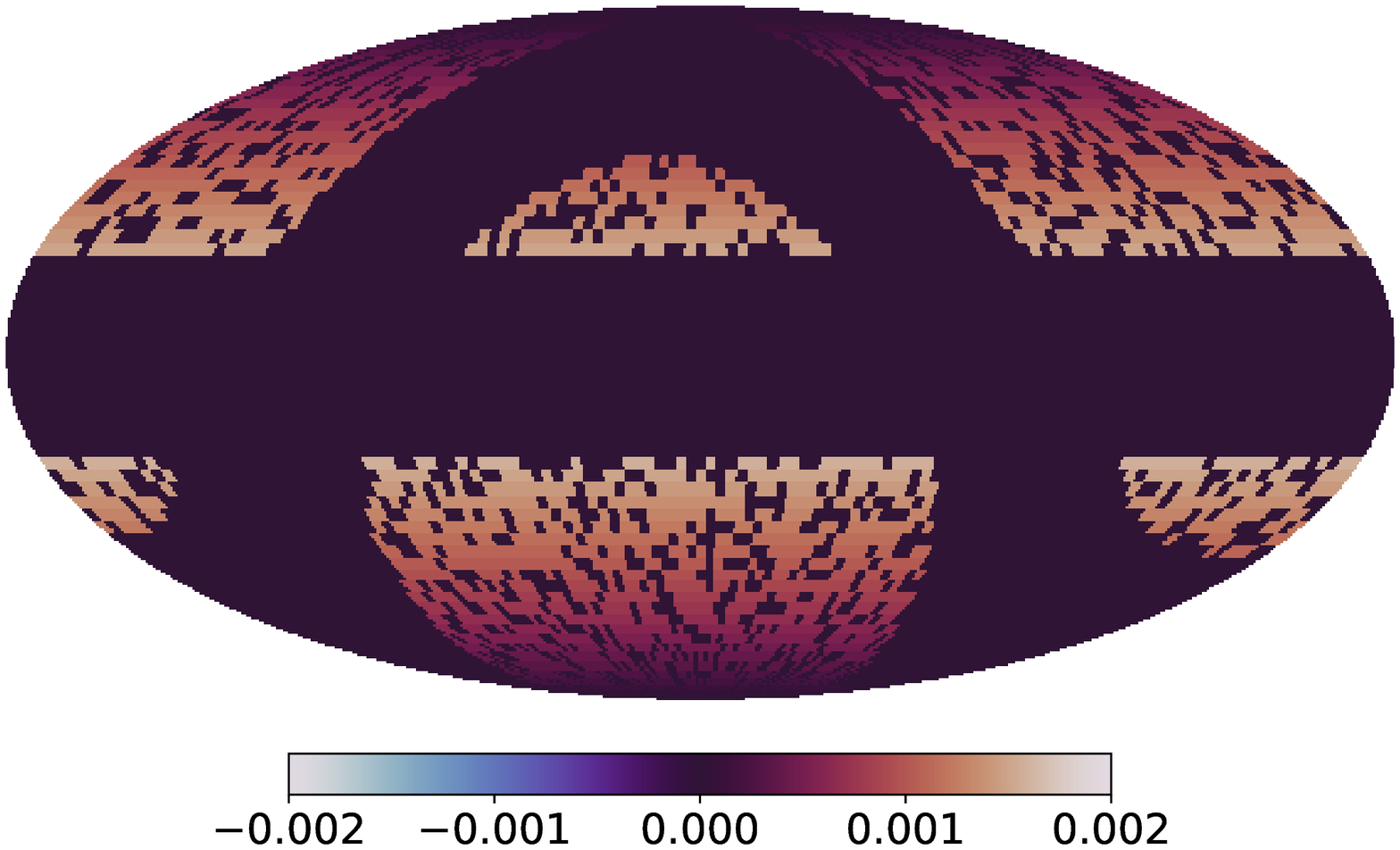}
\includegraphics[width=0.32\columnwidth]{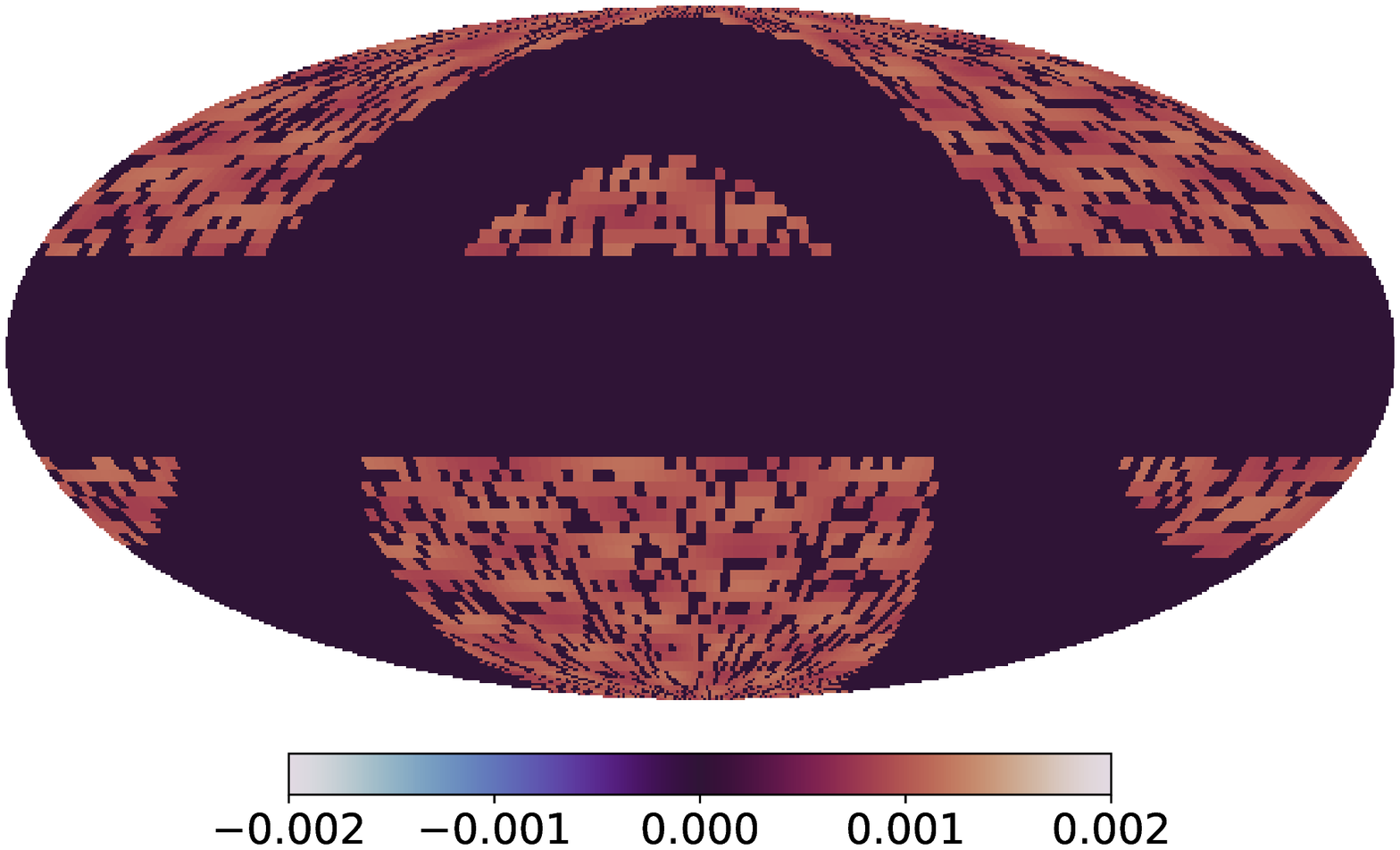}
\includegraphics[width=0.32\columnwidth]{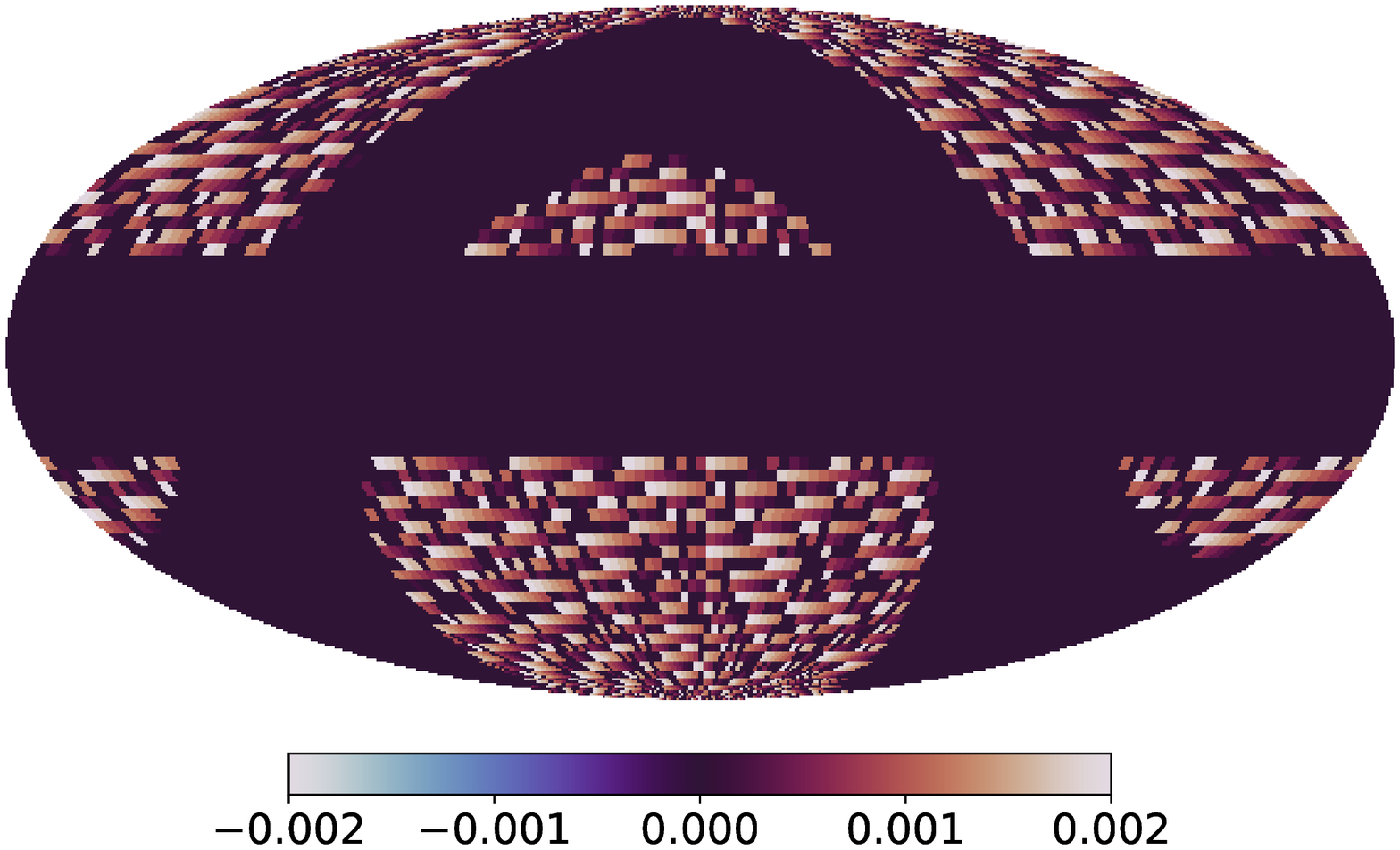}
\caption{The real part of the multiplicative bias field $m^R(\mathbf{\Omega})$, in the three simulated cases investigated. Cases 1, 2, 3 are shown left to right: simple galactic case, simple patch pattern and simple scanning pattern (see Section \ref{Simple Simulations}. Shown is a simulated celestial sphere, using a Mollweide Projection with $\theta=\phi=0$ at the North pole. The colour bar shows the amplitude of the biases, and the black regions show the regions where no data is present (i.e. the mask).}
\label{fig:fields}
\end{figure*} 

By combining equations (\ref{shtransform}) and (\ref{shtransform2}) we can find an expression that includes the effect of both biases and a mask 
\begin{eqnarray}
    \label{shtransform30}
    \widetilde\gamma^E_{\ell m}&=&
    \sum_{\ell'm'}
    \{
    \gamma^E_{\ell'm'}
     +\sum_{\ell''m''}
    [\gamma^E_{\ell''m''}({}^{m_0}W^+_{\ell\ell'm m'}+{}^{m_4}W^+_{\ell\ell'm m'})+
     \gamma^B_{\ell''m''}({}^{m_0}W^-_{\ell\ell'm m'}-{}^{m_4}W^-_{\ell\ell'm m'})]
     +c^E_{\ell''m''}\}{}^wW^+_{\ell\ell'm m'}\nonumber\\
     &+&
     \sum_{\ell'm'}\{\gamma^B_{\ell'm'}+\sum_{\ell''m''}
    [\gamma^B_{\ell''m''}({}^{m_0}W^+_{\ell\ell'm m'}-{}^{m_4}W^+_{\ell\ell'm m'})-
     \gamma^E_{\ell''m''}({}^{m_0}W^-_{\ell\ell'm m'}+{}^{m_4}W^-_{\ell\ell'm m'})]
     +c^B_{\ell' m'}\}{}^wW^-_{\ell\ell'm m'}\nonumber\\
    \widetilde\gamma^B_{\ell m}&=&      
    \sum_{\ell'm'}
    \{\gamma^B_{\ell'm'}+\sum_{\ell''m''}
    [\gamma^B_{\ell''m''}({}^{m_0}W^+_{\ell\ell'm m'}-{}^{m_4}W^+_{\ell\ell'm m'})-
     \gamma^E_{\ell''m''}({}^{m_0}W^-_{\ell\ell'm m'}+{}^{m_4}W^-_{\ell\ell'm m'})]
     +c^B_{\ell' m'}\}{}^wW^+_{\ell\ell'm m'}\nonumber\\
    &-&
    \sum_{\ell'm'}\{
    \gamma^E_{\ell'm'}
     +\sum_{\ell''m''}
    [\gamma^E_{\ell''m''}({}^{m_0}W^+_{\ell\ell'm m'}+{}^{m_4}W^+_{\ell\ell'm m'})+
     \gamma^B_{\ell''m''}({}^{m_0}W^-_{\ell\ell'm m'}-{}^{m_4}W^-_{\ell\ell'm m'})]
     +c^E_{\ell''m''}\}{}^wW^-_{\ell\ell'm m'}].
\end{eqnarray}
By comparing equations (\ref{shtransform}) and (\ref{shtransform30}) it can already be seen that the presence of a mask causes additional E and B-mode terms to occur in the power spectra. 

To simplify these expressions we note that the true BB field $C^{BB}_{\ell}\ll C^{EE}_{\ell}$. \cite{Schneider02} show that source redshift clustering can cause a small $B$-mode component, approximately three orders of magnitude less than the $E$-mode component over scales with $\ell\leq 5000$. Therefore the terms that contain multiplicative bias terms combined with the $B$-mode should be small i.e. we set terms $\mathcal{O}(m \gamma^B)=0$, but the unaffected $B$-mode component may be non-negligible. In this case these expressions simplify to 
\begin{eqnarray}
    \label{shtransform3}
    \widetilde\gamma^E_{\ell m}&=&
    \sum_{\ell'm'}
    \{
    \gamma^E_{\ell'm'}
     +\sum_{\ell''m''}
     \gamma^E_{\ell''m''}{}^{m_0+m_4}W^+_{\ell\ell'm m'}
     +c^E_{\ell''m''}\}{}^wW^+_{\ell\ell'm m'}
     +
     \sum_{\ell'm'}\{\gamma^B_{\ell'm'}-\sum_{\ell''m''}
     \gamma^E_{\ell''m''}{}^{m_0+m_4}W^-_{\ell\ell'm m'}
     +c^B_{\ell' m'}\}{}^wW^-_{\ell\ell'm m'}\nonumber\\
    \widetilde\gamma^B_{\ell m}&=&      
    \sum_{\ell'm'}
    \{\gamma^B_{\ell'm'}-\sum_{\ell''m''}
     \gamma^E_{\ell''m''}{}^{m_0+m_4}W^-_{\ell\ell'm m'}
     +c^B_{\ell' m'}\}{}^wW^+_{\ell\ell'm m'}
    -
    \sum_{\ell'm'}\{
    \gamma^E_{\ell'm'}
     +\sum_{\ell''m''}
    \gamma^E_{\ell''m''}{}^{m_0+m_4}W^+_{\ell\ell'm m'}
     +c^E_{\ell''m''}\}{}^wW^-_{\ell\ell'm m'}.
\end{eqnarray}
We combine the weight factors for the $m_0$ and $m_4$ terms and note that in this expression the real and imaginary parts of these fields propagate as sums i.e. one can write a total multiplicative bias field like $m^R(\mathbf{\Omega})=m_0^R(\mathbf{\Omega})+m_4^R(\mathbf{\Omega})$ and $m^I(\mathbf{\Omega})=m_0^I(\mathbf{\Omega})+m_4^I(\mathbf{\Omega})$.

\subsection{Power Spectra}
The power spectra estimates for the measured shear can now be computed by taking the correlation of the spherical harmonic coefficients from equation (\ref{shtransform3}) where 
\begin{eqnarray}
\label{eqsize}
\widetilde C^{GH}_{\ell,ij}&\equiv& \frac{1}{2\ell+1}\sum_m 
\widetilde\gamma^G_{\ell m,i}\widetilde\gamma^{H,*}_{\ell m,j}
\end{eqnarray}
for $G=(E, B)$ and $H=(E, B)$, where $i$ and $j$ are labels tomographic bins delineating galaxy populations defined by redshift or colour \cite{2019arXiv190106495K}.

We will assume that the true $EB$ and $BE$ power spectra are zero $C^{EB}_{\ell,ij}=C^{BE}_{\ell,ij}=0$, which should be the case in all but the most exotic dark energy models that cause parity-violating modes \citep{2013LRR....16....6A}. Given this assumption, the estimated $EE$ power spectra is given by  
\begin{eqnarray}
\label{full}
\widetilde C^{EE}_{\ell,ij}&=&{\sum_{\ell'}}
{\mathcal M}^{++}_{\ell\ell',ij} [C^{EE}_{\ell',ij}+C^{c_E c_E}_{\ell',ij} +C^{E c_E}_{\ell',ij}+C^{c_E E}_{\ell',ij}]+{\mathcal M}^{+-}_{\ell\ell',ij} [C^{E c_B}_{\ell',ij}+C^{c_E  B}_{\ell',ij}]\nonumber\\
&+&{\mathcal M}^{--}_{\ell\ell',ij} [C^{BB}_{\ell',ij}+C^{c_B c_B}_{\ell',ij}+C^{B c_B}_{\ell',ij}+C^{c_B B}_{\ell',ij}]+{\mathcal M}^{-+}_{\ell\ell',ij} [C^{c_B E}_{\ell',ij}+C^{B c_E}_{\ell',ij}]\nonumber\\
&+&(P^{+++}_{\ell\ell',ij}+P^{+++,*}_{\ell\ell',ij})C^{EE}_{\ell',ij}
-(P^{+--}_{\ell\ell',ij}+P^{+--,*}_{\ell\ell',ij})C^{EE}_{\ell',ij}\nonumber\\
&+&(T^{+++}_{\ell\ell',ij}+T^{+++,*}_{\ell\ell',ij})C^{c_EE}_{\ell',ij}
+(T^{--+}_{\ell\ell',ij}+T^{--+,*}_{\ell\ell',ij})C^{c_EE}_{\ell',ij}
+(T^{++-}_{\ell\ell',ij}+T^{++-,*}_{\ell\ell',ij})C^{Ec_B}_{\ell',ij}
+(T^{---}_{\ell\ell',ij}+T^{---,*}_{\ell\ell',ij})C^{Ec_B}_{\ell',ij}\nonumber\\
&+&(Q^{++++}_{\ell,ij}+Q^{----}_{\ell,ij})C^{EE}_{\ell',ij}
-(Q^{-++-}_{\ell,ij}+Q^{-++-,*}_{\ell,ij})C^{EE}_{\ell',ij}.
\end{eqnarray}
The various terms in the full expression are
\begin{eqnarray}
\label{matrices}
    {\mathcal M}^{XY}_{\ell\ell',ij}&=&\frac{1}{2\ell+1}\sum_{mm'} {}^wW^X_{\ell\ell'm m',i}{}^wW^{Y,*}_{\ell\ell' m m',j}\nonumber\\
    P^{XYZ}_{\ell\ell',ij}&=&\frac{1}{2\ell+1}\sum_{mm'}\sum_{\tilde\ell\tilde m}
    {}^wW^X_{\ell\ell'm m',i}{}^{m_0+m_4}W^{Y,*}_{\tilde\ell\ell'\tilde m m',j}{}^wW^{Z,*}_{\ell\tilde\ell m \tilde m,j}\nonumber\\
    T^{XYZ}_{\ell\ell',ij}&=&\frac{1}{2\ell+1}\sum_{mm'}\sum_{\tilde\ell\tilde m}
    {}^wW^X_{\ell\tilde\ell m m',i}
    {}^{m_0+m_4}W^{Y}_{\tilde\ell\ell' m'\tilde m,j}{}^wW^{Z,*}_{\ell\ell' m \tilde m,j}\nonumber\\
    Q^{WXYZ}_{\ell\ell',ij}&=&
    \frac{1}{2\ell+1}\sum_{mm'}\sum_{\tilde\ell\tilde m}
    \sum_{\tilde\ell'\tilde m'}
    {}^{m_0+m_4}W^W_{\ell\tilde\ell' \tilde m m',i}
    {}^wW^X_{\ell\tilde\ell m \tilde m,i}{}^{m_0+m_4}W^{Y,*}_{\tilde\ell'\ell' \tilde m' m',j}
    {}^wW^{Z,*}_{\ell\tilde\ell' m \tilde m',j}
\end{eqnarray}
where $W=(+,-)$, $X=(+,-)$, $Y=(+,-)$, $Z=(+,-)$. Power spectra in equation (\ref{full})  are labelled in their superscripts e.g. $EE$, $EB$, $BB$ or $c_E c_E$,  $c_E c_B$, $c_B c_B$ for additive bias terms. On the left hand side of equation (\ref{full}) we define the measured power spectrum and compare this to the power spectrum that would have been measured in the absence of systematic effects. We note however that terms in the window functions (${}^{m_0+m_4}W^X_{\ell\ell'm m'}$ and ${}^wW^X_{\ell\ell'm m'}$) are derived via the ensemble-average of equation (\ref{eqsize}), and make use of the statistical rotational invariance of the ensemble-averaged harmonic modes. Therefore equation (\ref{full}) is a hybrid of ensemble-averaged terms and terms that are not averaged which may be non-zero only for a given realisation. This is tested numerically in Section \ref{Simple Simulations}. We do not present the EB and BB equivalents here since, as demonstrated later in the paper a linear decoupled-field expression is sufficient to characterise the impact of biases. Whilst equation (\ref{full}) gives the full tomographic expression, for the remainder of this paper we will only consider a non-tomographic case for simplicity i.e. $i=j$.
\\

\subsection{Linear Decoupled expressions}
\label{Linearised Terms}
Here we simplify the analysis by exploring two assumptions. The first is a \emph{linearity} assumption that terms of order $m^2$, $c^2$ or $mc$ and higher are negligible. The second is a \emph{decoupled} assumption that the spherical harmonic transform of the mask has no correlation with the spherical harmonic transform of the multiplicative bias field. 

In comparison with \cite{K19} we can identify the $P$ terms to be similar to the linear multiplicative bias terms in that paper, which were shown to only depend on the mean of the multiplicative bias field. In the case of masks these linear terms do not in general reduce to the mean of the multiplicative bias field since the mask may be coupled to the multiplicative bias field; this is something we numerically investigate in Section \ref{Simple Simulations}. However, if the multiplicative bias field is constant and/or not strongly coupled to the mask, then 
\begin{eqnarray}
\label{dec}
    P^{X+Z}_{\ell\ell'}&=&\langle m^R(\mathbf{\Omega})\rangle {\mathcal M}^{XZ}_{\ell\ell'}\nonumber\\
    P^{X-Z}_{\ell\ell'}&=&-\langle m^I(\mathbf{\Omega})\rangle {\mathcal M}^{XZ}_{\ell\ell'}
\end{eqnarray}
where $\langle m^R(\mathbf{\Omega})\rangle=\langle m_0^R(\mathbf{\Omega})\rangle+\langle m_4^R(\mathbf{\Omega})\rangle$ and $\langle m^I(\mathbf{\Omega})\rangle=\langle m_0^I(\mathbf{\Omega})\rangle+\langle m_4^I(\mathbf{\Omega})\rangle$ are the mean of the real and imaginary parts of the sum of the multiplicative bias fields respectively. 

Assuming no coupling and to linear order in biases we find that 
\begin{eqnarray}
\label{lindec}
\widetilde C^{EE}_{\ell}&\approx&{\sum_{\ell'}}
{\mathcal M}^{++}_{\ell\ell'} [(1+2\langle m^R(\mathbf{\Omega})\rangle)C^{EE}_{\ell'}+2C^{E c_E}_{\ell'}]
+{\mathcal M}^{--}_{\ell\ell'} [C^{BB}_{\ell'}+2C^{B c_B}_{\ell'}]\nonumber\\
\widetilde C^{BB}_{\ell}&\approx&{\sum_{\ell'}}
{\mathcal M}^{--}_{\ell\ell'} [(1+2\langle m^R(\mathbf{\Omega})\rangle)C^{EE}_{\ell'}+2C^{E c_E}_{\ell'}]
+{\mathcal M}^{++}_{\ell\ell'} [C^{BB}_{\ell'}+2C^{B c_B}_{\ell'}]\nonumber\\
\widetilde C^{EB}_{\ell}&\approx&{\sum_{\ell'}}
{\mathcal M}^{++}_{\ell\ell'} [C^{E c_B}_{\ell'}+C^{c_EB}_{\ell'}-\langle m^I(\mathbf{\Omega})\rangle C^{EE}_{\ell'}]
+{\mathcal M}^{--}_{\ell\ell'} [C^{c_BE}_{\ell'}-C^{B c_E}_{\ell'}+\langle m^I(\mathbf{\Omega})\rangle C^{EE}_{\ell'}]\nonumber\\
&+&{\mathcal M}^{+-}_{\ell\ell'} [(1+2\langle m^R(\mathbf{\Omega})\rangle)(C^{EE}_{\ell'}+2C^{E c_E}_{\ell'})]
+{\mathcal M}^{-+}_{\ell\ell'} [C^{BB}_{\ell'}+2C^{B c_B}_{\ell'}]\nonumber\\
\widetilde C^{BE}_{\ell}&\approx&{\sum_{\ell'}}
{\mathcal M}^{++}_{\ell\ell'} [C^{E c_B}_{\ell'}+C^{c_EB}_{\ell'}-\langle m^I(\mathbf{\Omega})\rangle C^{EE}_{\ell'}]
+{\mathcal M}^{--}_{\ell\ell'} [C^{c_BE}_{\ell'}-C^{B c_E}_{\ell'}+\langle m^I(\mathbf{\Omega} \rangle C^{EE}_{\ell'}]\nonumber\\
&+&{\mathcal M}^{-+}_{\ell\ell'} [(1+2\langle m^R(\mathbf{\Omega})\rangle)(C^{EE}_{\ell'}+2C^{E c_E}_{\ell'})]
+{\mathcal M}^{+-}_{\ell\ell'} [C^{BB}_{\ell'}+2C^{B c_B}_{\ell'}].
\end{eqnarray}
where we assume that $C^{E c_E}_{\ell'}=C^{c_E E}_{\ell'}$ and similar for other times (i.e. a non-tomographic case), and we note that ${\mathcal M}^{+-}_{\ell\ell'}+{\mathcal M}^{-+}_{\ell\ell'}=0$.

\subsection{Unlensed random ellipticity contribution}
To include the effect of the stochastic ellipticity field (i.e. the random uncorrelated and unlensed ellipticities of galaxies) in the expressions above we add a term to the true shear $\gamma(\mathbf{\Omega})+n(\mathbf{\Omega})$. We note that the presence of a multiplicative bias in the measurement will affect the observed stochastic ellipticity component  
\begin{eqnarray}
    \label{noise1}
    \widetilde\gamma(\mathbf{\Omega})&=&W(\mathbf{\Omega})\{[1+m_0(\mathbf{\Omega})][\gamma(\mathbf{\Omega})+n(\mathbf{\Omega})]+m_4(\mathbf{\Omega})[\gamma^*(\mathbf{\Omega})+n^*(\mathbf{\Omega})]+c(\mathbf{\Omega})\},
\end{eqnarray}
here $n(\mathbf{\Omega})$ is the true underlying uncorrelated galaxy ellipticity. This contribution is the zero-lag intrinsic ellipticity field \citep{2001ApJ...559..552C,2016MNRAS.461.4343L,2015JCAP...08..015B}, which in the case of a finite number of galaxies is expressed as a shot noise term; see \cite{2019PhRvD.100j3506B} for a discussion. 

The shot noise component of the uncorrelated ellipticity term for a finite number of galaxies in a sample has the properties that 
\begin{eqnarray}
\label{noiseq}
n^E_{\ell m}&=&n^B_{\ell m}\nonumber\\
N_{\ell}&=&\langle n^X_{\ell m}n^{Y,*}_{\ell m}\rangle=\frac{\delta^K_{XY}\delta^K_{\ell\ell}\delta^K_{mm}\sigma_e^2}{N_{\rm gal}}\nonumber\\
N_{\ell}&=&\langle n^X_{\ell m}\gamma^{Y,*}_{\ell m}\rangle=0
\end{eqnarray}
where $\sigma_e^2$ is the intrinsic (unlensed) variance of the ellipticities, $N_{\rm gal}$ is the effective number of galaxies in the observations \citep[for a discussion of the effective number density see][]{2019PhRvD.100j3506B, 2013MNRAS.434.2121C}, and $X$ and $Y$ are $(E,B)$. We note that any additional noise caused by the measurement process itself (e.g. sky noise, detector noise etc.) is already captured in a stochastic contribution to the $c(\mathbf{\Omega})$ term. 

In this case we have a general expression that is 
\begin{eqnarray}
\label{gencase3}
\widetilde C^{EE}_{\ell}&\approx&{\sum_{\ell'}}
{\mathcal M}^{++}_{\ell\ell'} [(1+2\langle m^R(\mathbf{\Omega})\rangle)(C^{EE}_{\ell'}+N_{\ell'})+2C^{E c_E}_{\ell'}]
+{\mathcal M}^{--}_{\ell\ell'} [C^{BB}_{\ell'}+(1+2\langle m^R(\mathbf{\Omega})\rangle)N_{\ell'}+2C^{B c_B}_{\ell'}]\nonumber\\
\widetilde C^{BB}_{\ell}&\approx&{\sum_{\ell'}}
{\mathcal M}^{--}_{\ell\ell'} [(1+2\langle m^R(\mathbf{\Omega})\rangle)(C^{EE}_{\ell'}+N_{\ell'})+2C^{E c_E}_{\ell'}]
+{\mathcal M}^{++}_{\ell\ell'} [C^{BB}_{\ell'}+(1+2\langle m^R(\mathbf{\Omega})\rangle)N_{\ell'}+2C^{B c_B}_{\ell'}]\nonumber\\
\widetilde C^{EB}_{\ell}&\approx&{\sum_{\ell'}}
{\mathcal M}^{++}_{\ell\ell'} [C^{E c_B}_{\ell'}+C^{c_EB}_{\ell'}-\langle m^I(\mathbf{\Omega})\rangle (C^{EE}_{\ell'}+N_{\ell'})]
+{\mathcal M}^{--}_{\ell\ell'} [C^{c_BE}_{\ell'}-C^{B c_E}_{\ell'}+\langle m^I(\mathbf{\Omega})\rangle (C^{EE}_{\ell'}+N_{\ell'})]\nonumber\\
&+&{\mathcal M}^{+-}_{\ell\ell'} [(1+2\langle m^R(\mathbf{\Omega})\rangle)(C^{EE}_{\ell'}+N_{\ell'}+2C^{E c_E}_{\ell'})]
+{\mathcal M}^{-+}_{\ell\ell'} [C^{BB}_{\ell'}+(1+2\langle m^R(\mathbf{\Omega})\rangle)N_{\ell'}+2C^{B c_B}_{\ell'}]\nonumber\\
\widetilde C^{BE}_{\ell}&\approx&{\sum_{\ell'}}
{\mathcal M}^{++}_{\ell\ell'} [C^{E c_B}_{\ell'}+C^{c_EB}_{\ell'}-\langle m^I(\mathbf{\Omega})\rangle (C^{EE}_{\ell'}+N_{\ell'})]
+{\mathcal M}^{--}_{\ell\ell'} [C^{c_BE}_{\ell'}-C^{B c_E}_{\ell'}+\langle m^I(\mathbf{\Omega} \rangle (C^{EE}_{\ell'}+N_{\ell'})]\nonumber\\
&+&{\mathcal M}^{-+}_{\ell\ell'} [(1+2\langle m^R(\mathbf{\Omega})\rangle)(C^{EE}_{\ell'}+N_{\ell'}+2C^{E c_E}_{\ell'})]
+{\mathcal M}^{+-}_{\ell\ell'} [C^{BB}_{\ell'}+(1+2\langle m^R(\mathbf{\Omega})\rangle)N_{\ell'}+2C^{B c_B}_{\ell'}].
\end{eqnarray}
We note that the noise term adding to the BB part is multiplied by $(1+2\langle m^R(\mathbf{\Omega})\rangle)$, but recall that we have assumed that terms that contain multiplicative bias terms combined with the true  $B$-mode should be small i.e. $\mathcal{O}(m C^{BB}_{\ell})=0$.

\subsection{Power spectrum combinations}
\label{pcombosec}
Equation (\ref{gencase3}) is the most general case, however to simplify further one can make several reasonable assumptions. The first is that there is no true BB field $C^{BB}_{\ell}=0$ which should be a good approximation; however we reiterate that \cite{Schneider02} show that source redshift clustering can cause a small $B$-mode component. The second is that the correlation between the additive bias and the shear field is small, which given that the majority of additive biases have a source in instrumental or optical effects, is a reasonable assumption. 

We apply these approximations to the EE and BB cases, and we take some combinations of power spectra to highlight the inter-relationships between them, 
\begin{eqnarray}
\widetilde C^{EE}_{\ell}&\approx&[1+2\langle m^R(\mathbf{\Omega})\rangle]{\sum_{\ell'}}
\{{\mathcal M}^{++}_{\ell\ell'} [(C^{EE}_{\ell'}+N_{\ell'})]
+{\mathcal M}^{--}_{\ell\ell'} N_{\ell'}\}\nonumber\\
\widetilde C^{BB}_{\ell}&\approx&[1+2\langle m^R(\mathbf{\Omega})\rangle]{\sum_{\ell'}}
\{{\mathcal M}^{--}_{\ell\ell'} [(C^{EE}_{\ell'}+N_{\ell'})]
+{\mathcal M}^{++}_{\ell\ell'} N_{\ell'}\}\nonumber\\
\widetilde C^{EE}_{\ell}+\widetilde C^{BB}_{\ell}&\approx&[1+2\langle m^R(\mathbf{\Omega})\rangle]\sum_{\ell'}[{\mathcal M}^{++}_{\ell\ell'}+{\mathcal M}^{--}_{\ell\ell'}][C^{EE}_{\ell'}+2N_{\ell'}]\nonumber\\
\widetilde C^{EE}_{\ell}-\widetilde C^{BB}_{\ell}&\approx&[1+2\langle m^R(\mathbf{\Omega})\rangle]\sum_{\ell'}[{\mathcal M}^{++}_{\ell\ell'}-{\mathcal M}^{--}_{\ell\ell'}]C^{EE}_{\ell'}\nonumber\\
\widetilde C^{EB}_{\ell}+\widetilde C^{BE}_{\ell}&\approx&-2\langle m^I(\mathbf{\Omega})\rangle\sum_{\ell'}[{\mathcal M}^{++}_{\ell\ell'}-{\mathcal M}^{--}_{\ell\ell'}]C^{EE}_{\ell'}
+[1+2\langle m^R(\mathbf{\Omega})\rangle]{\sum_{\ell'}}
[{\mathcal M}^{+-}_{\ell\ell'}+{\mathcal M}^{-+}_{\ell\ell'}][C^{EE}_{\ell'}+2N_{\ell'}]\nonumber\\
\widetilde C^{EB}_{\ell}-\widetilde C^{BE}_{\ell}&\approx&[1+2\langle m^R(\mathbf{\Omega})\rangle]\sum_{\ell'}[{\mathcal M}^{+-}_{\ell\ell'}-{\mathcal M}^{-+}_{\ell\ell'}]C^{EE}_{\ell'},
\end{eqnarray}
where we have chosen the combinations that highlight the interrelations clearly. We note that in the case that there is no mask ${\mathcal M}^{++}_{\ell\ell'}=\delta^K_{\ell\ell'}$ and ${\mathcal M}^{--}_{\ell\ell'}={\mathcal M}^{+-}_{\ell\ell'}={\mathcal M}^{-+}_{\ell\ell'}=0$ these expressions reduce to the unmasked case given in \cite{K19} (equation 12). Finally we note that in general ${\mathcal M}^{+-}_{\ell\ell'}\approx 0$ and ${\mathcal M}^{-+}_{\ell\ell'}\approx 0$.
\begin{figure*}
\centering
\includegraphics[width=0.32\columnwidth]{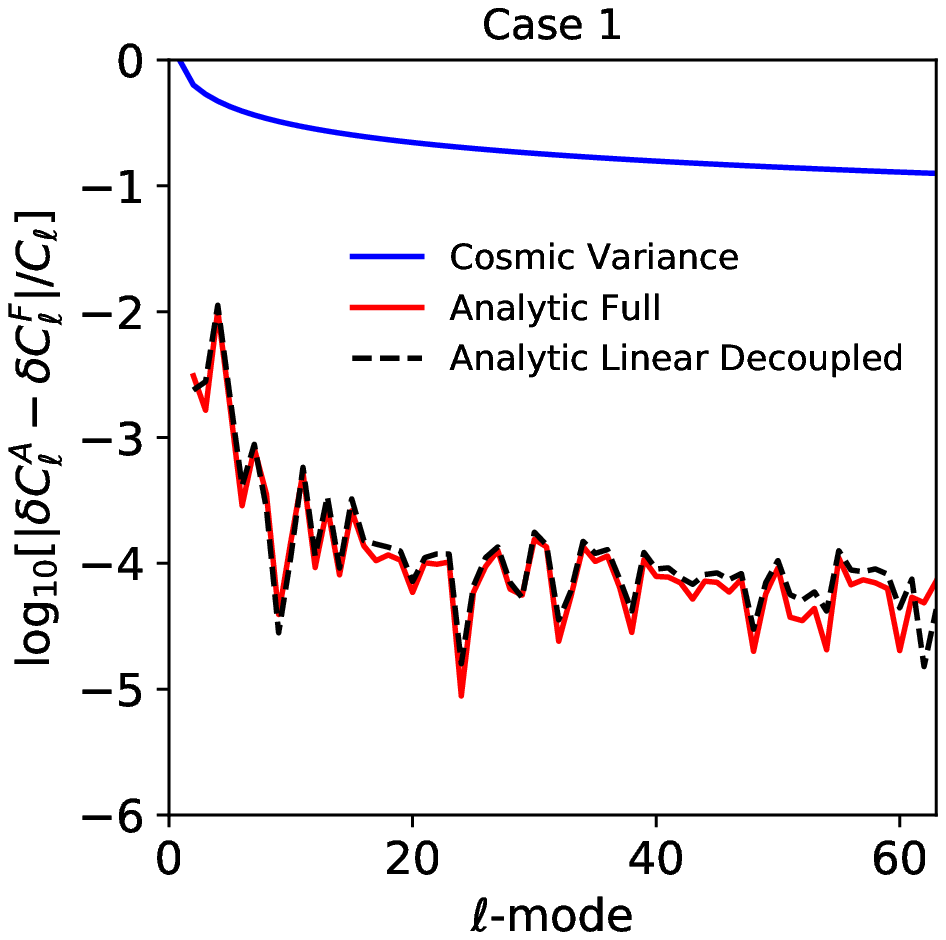}
\includegraphics[width=0.32\columnwidth]{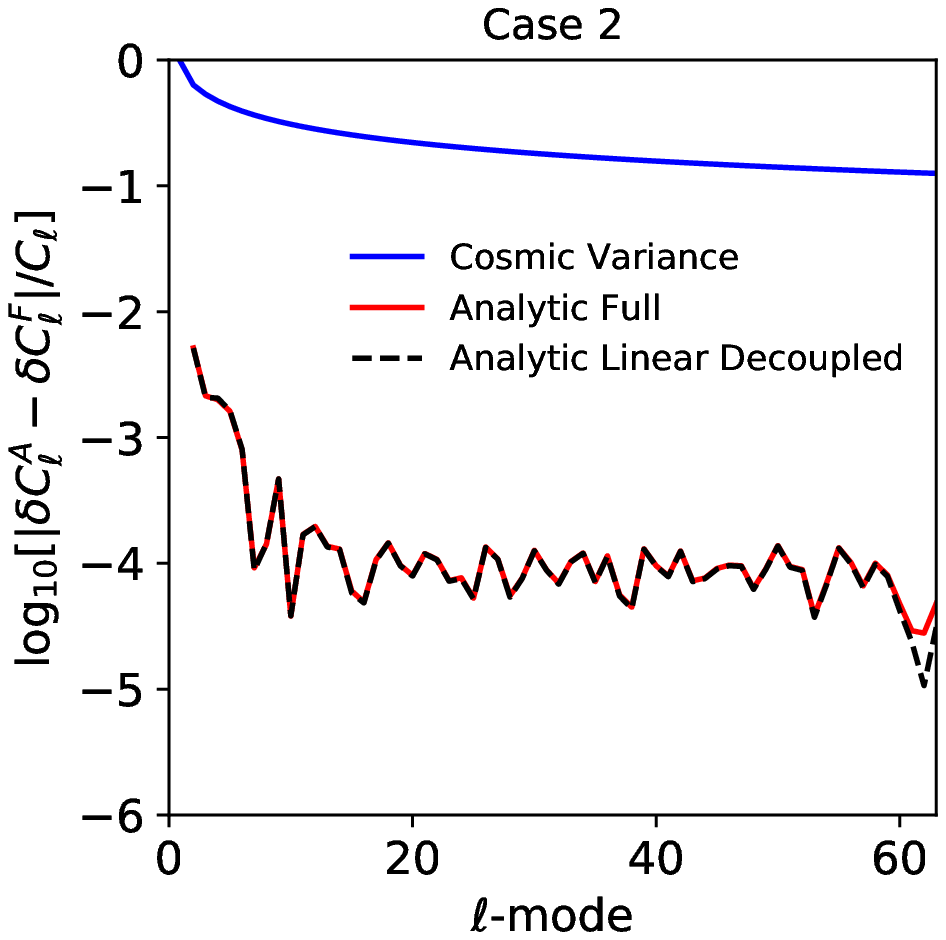}
\includegraphics[width=0.32\columnwidth]{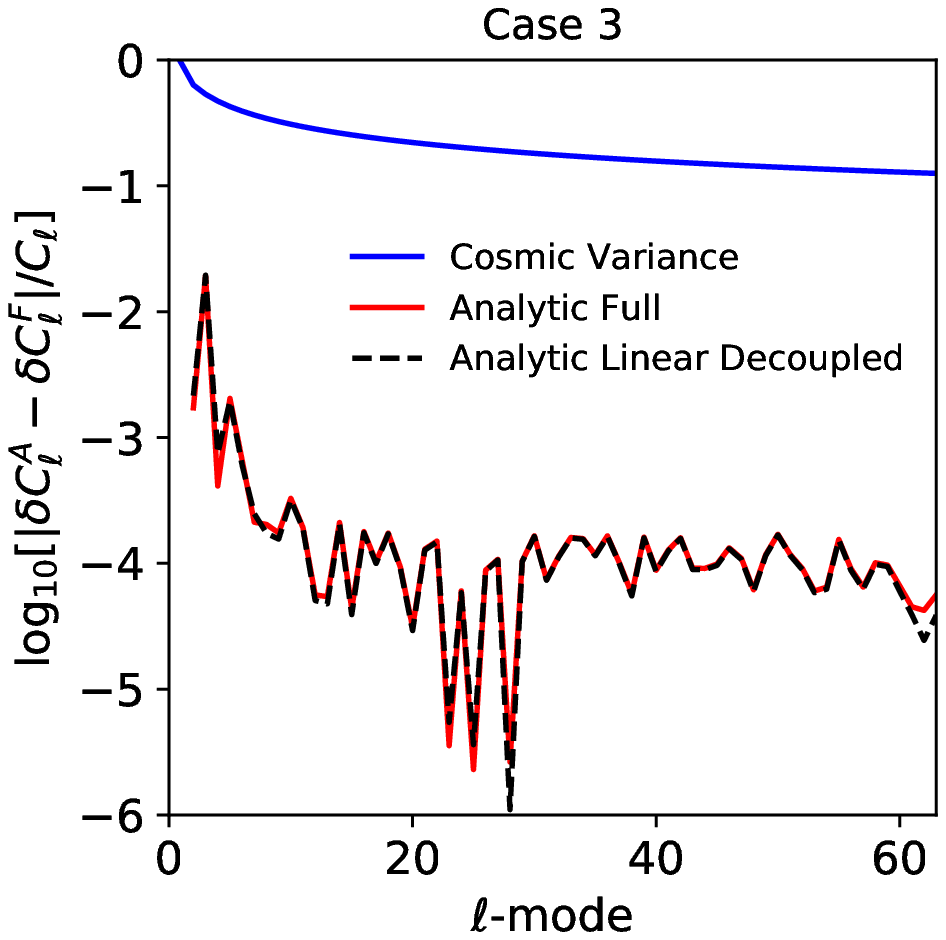}
\includegraphics[width=0.33\columnwidth]{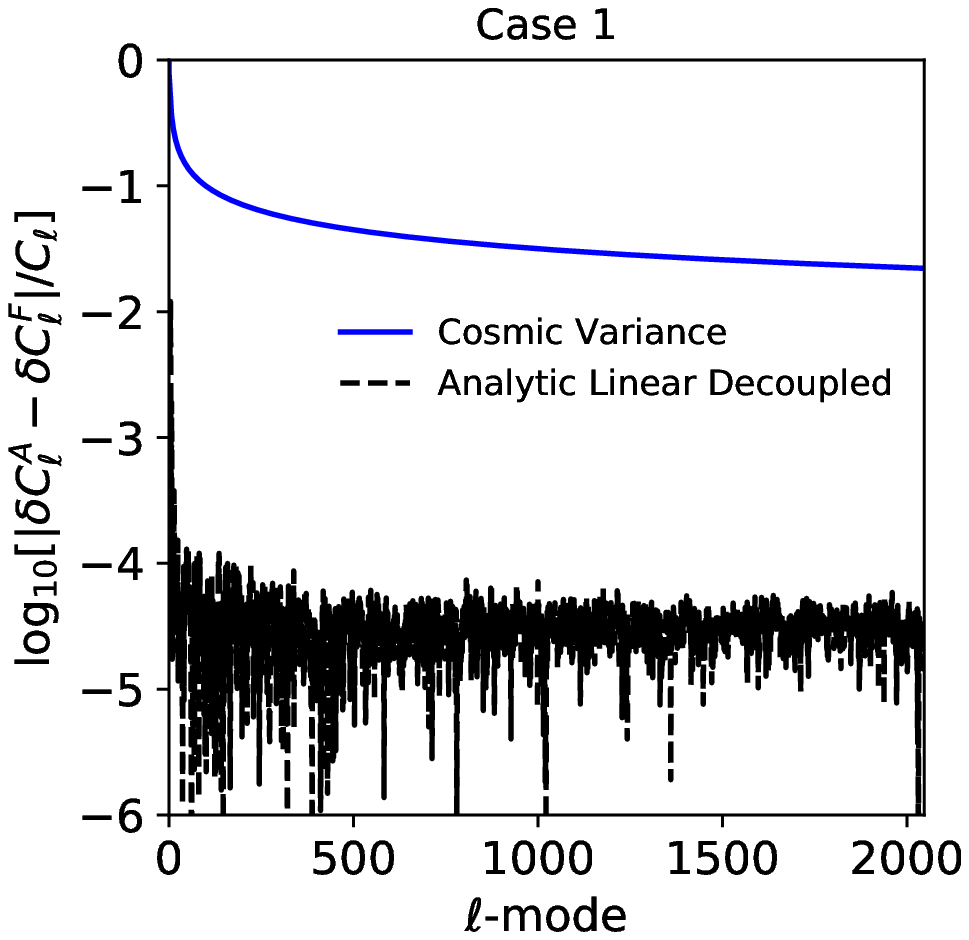}
\includegraphics[width=0.33\columnwidth]{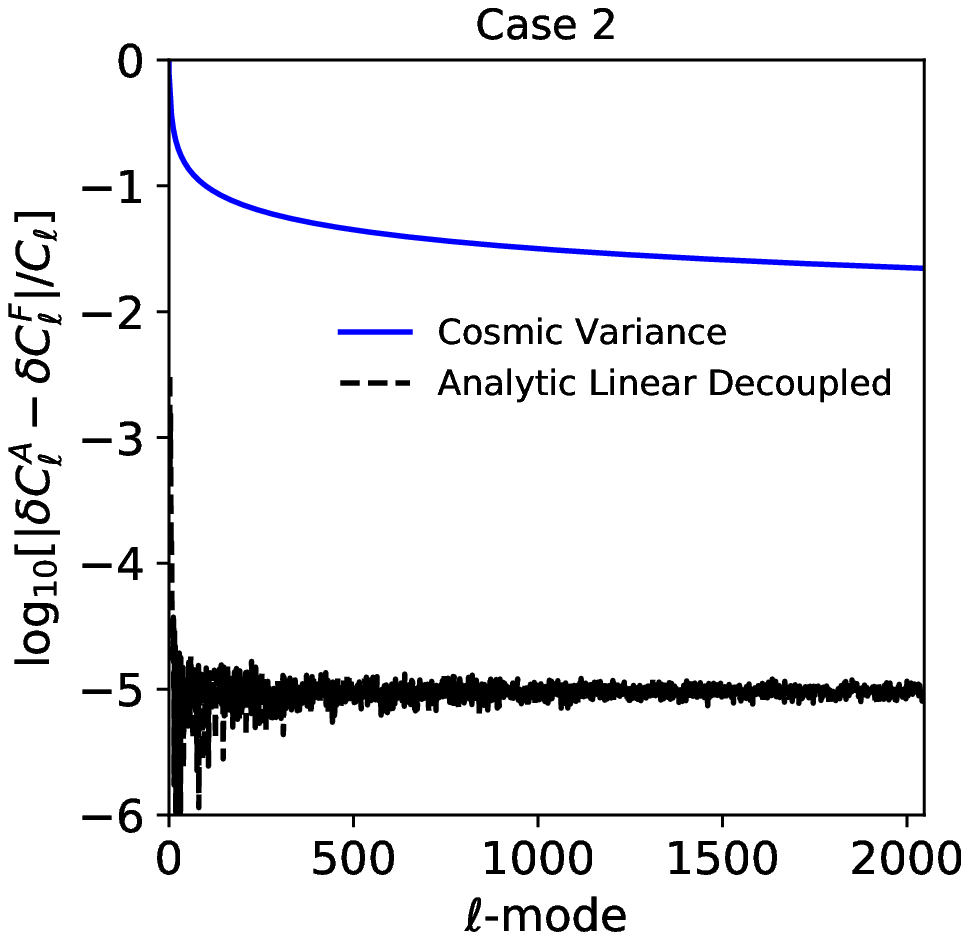}
\includegraphics[width=0.33\columnwidth]{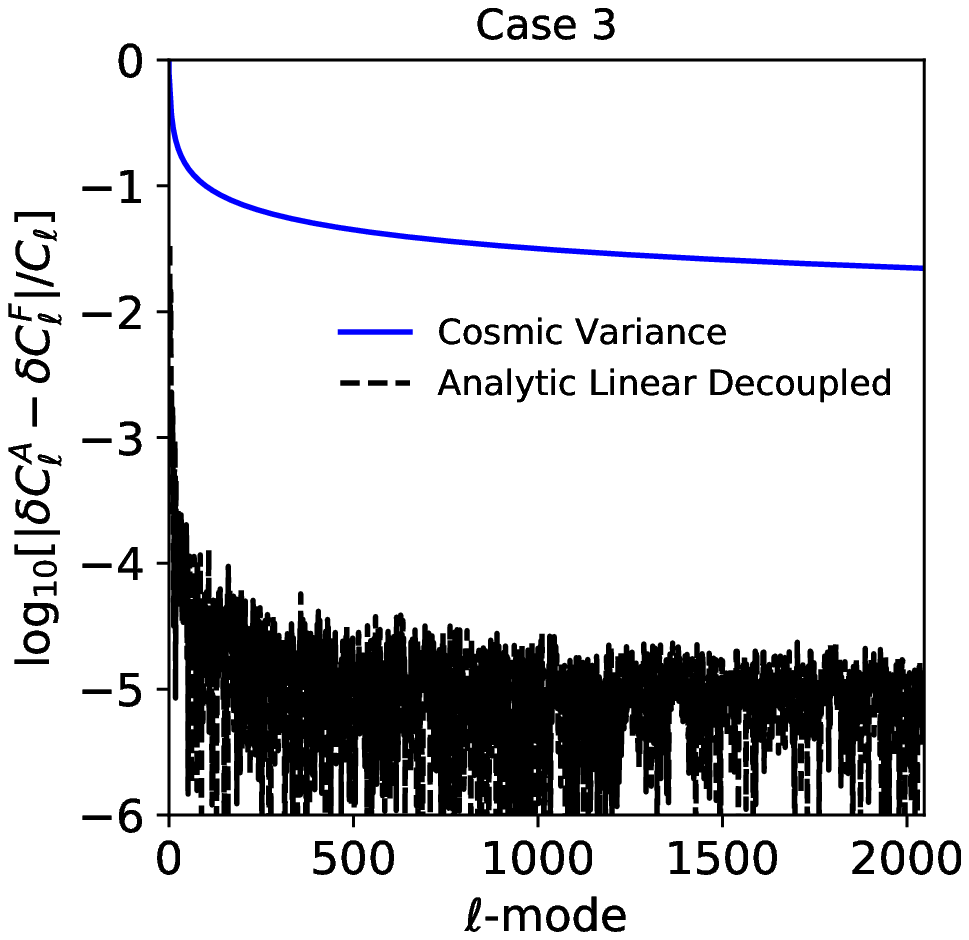}
\caption{The fractional difference between the residual power spectrum calculated analytically $\delta C^A_{\ell}$ and that found using a forward model $\delta C^F_{\ell}$ for the three cases considered. The fractional error is with respect to the input cosmic shear power spectrum. For the analytic case we compute the full expression (using equation \ref{full}), and using the linear decoupled approximation (equation \ref{lindec}). We plot the cosmic variance error on the cosmic shear power spectrum for comparison. In upper panels we include both calculations for $L=64$ (limited due to the complexity of the calculations in the full expression), in the lower panels we include only the linear decoupled approximation for $L=2048$. }
\label{fig:power}
\end{figure*}

\subsection{Discussion}
There are many combinations of cosmic shear power spectra that can be made each of which will depend on the multiplicative bias and the stochastic variance of the ellipticity field in a different way, both of which are unknown quantities. The most commonly used approached is to use the EE only power spectrum and use the BB power as a consistency test of the level of systematic effects in the data. However, as we have shown the BB power contains information on the multiplicative bias via the observed stochastic ellipticity component. Therefore one can construct a joint EE and BB likelihood where the likelihoods of the EE and BB cases would be summed to form a combined likelihood. A third approach is to subtract the BB from the EE power to form EE-BB which will be dependent on $m$ but not $\sigma_e$. We summarise these in Table \ref{stats}. These combinations are applicable even after deconvolving the mask (mask deconvolution is a separate point compared to the fact that BB power provides information on the multiplicative bias). 
\begin{table*}
\begin{center}
\begin{tabular}{|c|c|c|c|}
\hline
Statistic&Observables&Model&Variance$/{\mathcal V}^2$\\
\hline
&&&\\
EE Only&$\widetilde C^{EE}_{\ell}$&$(1+2m)[C^{EE}_{\ell}+N_{\ell}]$&$(C^{EE}_{\ell})^2+N^2_{\ell}$\\
&&&\\
\hline
&&&\\
EE-BB&$\widetilde C^{EE}_{\ell}-\widetilde C^{BB}_{\ell}$&$(1+2m)[C^{EE}_{\ell}]$&$(C^{EE}_{\ell})^2+2N^2_{\ell}$\\
&&&\\
\hline
&&&\\
EE and BB (joint likelihood)&$\widetilde C^{EE}_{\ell}$, $\widetilde C^{BB}_{\ell}$&$(1+2m)[C^{EE}_{\ell}+N_{\ell}]$,$(1+2m)N_{\ell}$&$(C^{EE}_{\ell})^2+N^2_{\ell}$, $N^2_{\ell}$\\
&&&\\
\hline
\end{tabular}
\caption{A summary of the various ways that the EE and BB power spectra can be combined. This is a summary of equations (\ref{lindec}) in the all-sky case. The Gaussian part the variance on each estimator is shown in the all-sky case is shown (normalised by the cosmic variance, equation \ref{cv}) which is used in the Gaussian random field simulations in Section \ref{Simple Simulations}.} 
\label{stats}
\end{center}
\end{table*}

Since these statistics depend on unknown parameters $m$ and $\sigma_e$ these parameters need to be marginalised over, and the degeneracy with cosmological parameters will vary between the statistics. We note that marginalisation will always need to be performed in a final likelihood analysis since at best calibration simulations will provide calibration of $m$ with some uncertainty. To mitigate degeneracies, and as may be available from previous simulation/calibration data, one should apply a prior to these parameters. In Appendix A we show that the estimation of a cosmic shear amplitude $A$ will be biased by imposing a prior on $m$. We investigate these degeneracies and the impact of priors numerically in Section \ref{Multiplicative bias tests}. 

Throughout we do not attempt to estimate the true power spectrum via inversion of the mixing matrices. This is because when a large fraction of the sky is masked some modes are not observable (i.e. they are in the mask) leading to singular mixing matrices. The standard approach to mitigating this effect is to use band-powers, but such an approach is leads to a loss of information \citep{2002ApJ...567....2H}. 

\section{Simple Simulations}
\label{Simple Simulations}
In this Section we use simple simulations to test whether a linear multiplicative bias assumption is applicable, i.e. that higher order terms $\mathcal{O}(m^2)$ can be ignored, and whether the linear decoupled assumptions are reasonable (equation \ref{lindec}); and also to investigate marginalisation over an unknown residual multiplicative bias and ellipticity variance.

We use the same extreme multiplicative shear fields used to test the full-sky formalism in \cite{K19}, except with $c(\mathbf{\Omega})=0$ and $m^I(\mathbf{\Omega})=0$ which as discussed in \cite{K19} are reasonable approximations. The cases we consider are shown below. Note that we express these in terms of an arbitrary amplitude $\alpha$ since these are all normalised to have $\langle m^R(\mathbf{\Omega})\rangle = 2\times 10^{-3}$. The cases are: 
\begin{itemize}
    \item Case1 : Simple Galactic Plane,  $m^R(\mathbf{\Omega})=\alpha[\pi-|\phi-\pi|]$
    \item Case 2: Simple Patch Pattern, $m^R(\mathbf{\Omega})=10\alpha\sin(100|\phi-\pi|)\sin(100|\theta-\pi|)$, 
    \item Case 3: Simple Scanning Pattern,  $m^R(\mathbf{\Omega})=\alpha i$, where i is an iterative pixel number count, which is reset when $i=10$.
\end{itemize}
We use a mask that removes data from less than $20^{\circ}$ in both the galactic and ecliptic planes; and also $20\%$ of pixels at random, to represent an all sky-like mask with random patches removed -- this gives a total observed sky fraction of $f_{\rm sky}=0.4$. We show the masked bias fields in Figure \ref{fig:fields}.
\begin{figure*}
\centering
\includegraphics[width=0.33\columnwidth]{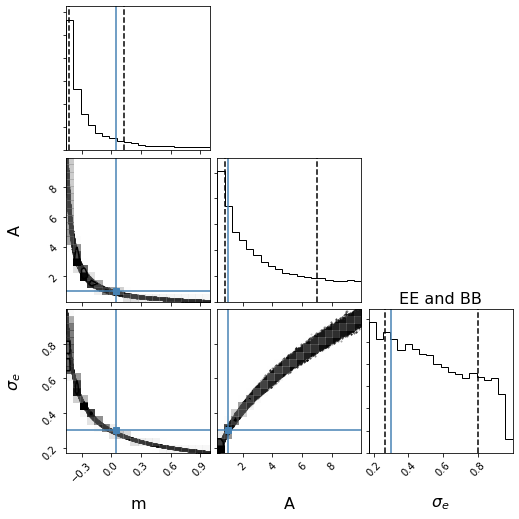}
\includegraphics[width=0.33\columnwidth]{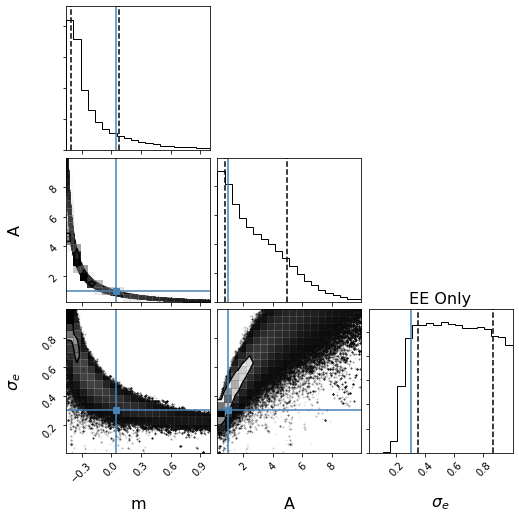}
\includegraphics[width=0.24\columnwidth]{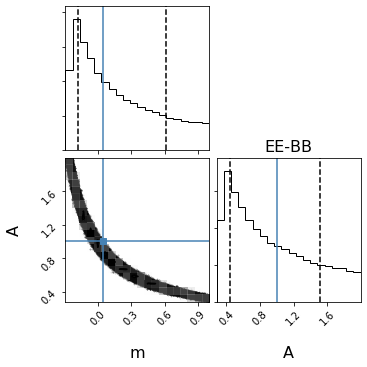}
\caption{The constraints on $m$, $A$ and $\sigma_e$ for Gaussian random field simulations described in Section \ref{Multiplicative bias tests} for those statistics in Table \ref{stats}. In this case the contours are $1,2,3$-$\sigma$ and the gray shading shows the density of the MCMC points. In this case only flat priors on all parameters are used with $-1\leq m\leq 1$, $0\leq A\leq 2$ and $0\leq \sigma_e\leq 1$. The blue lines show the input values and the dashed lines in the 1D histograms show the $1$-$\sigma$ errors about the median.}
\label{G2}
\end{figure*}

We compute\footnote{We use the {\tt massmappy} code \citep{wallis}, {\tt SSHT} \cite{ssht}, and sample the sphere using the sampling scheme of \cite{mw}.} the original $\gamma(\mathbf{\Omega})$ field using a Gaussian random field using a {\it Planck} $\Lambda$CDM cosmology \citep{planck}. The theoretical EE power spectrum, subject to the Limber \citep{1953ApJ...117..134L,2017MNRAS.469.2737K,2017JCAP...05..014L}, flat-sky \citep{10.1046/j.1365-8711.1998.02054.x}, flat-universe \citep{2018PhRvD..98b3522T}, prefactor-unity \citep{2017MNRAS.469.2737K} and reduced shear \citep{ad2} approximations is:
\begin{equation}
    \label{eq:cltheory}
    C^{EE}_{\ell} = \int_0^{\chi_{\rm H}} {\rm d}\chi \frac{q^2(\chi)}{\chi^2} P_{\delta} \left(\frac{\ell + 1/2}{\chi}, \chi\right),
\end{equation}
where $\chi$ is the comoving distance, $\chi_{\rm H}$ is the comoving distance to the horizon, $P_{\delta}$ is the matter power spectrum, and $q$ is the lensing kernel:
\begin{equation}
    \label{eq:lenskern}
    q(\chi) = \frac{3}{2}\Omega_{\rm M} \frac{H^2_0}{c^2} \frac{\chi}{a(\chi)} \int^{\chi_{\rm H}}_\chi {\rm d}\chi'\, n(\chi')\, \frac{\chi'-\chi}{\chi},
\end{equation}
where $\Omega_{\rm M}$ is the present-day dimensionless total matter density of the Universe, $H_0$ is the Hubble constant, $c$ is the speed of light in a vacuum, $a$ is the scale factor of the Universe, and $n(\chi)$ is the galaxy distribution function of the survey. In this work, we use the photometric DES Year 1 galaxy distribution\footnote{Data available at \url{http://desdr-server.ncsa.illinois.edu/despublic/y1a1\_files/redshift\_bins/}} \citep{des2}. The matter power spectrum is calculated using the publicly available \texttt{CAMB} cosmology package \citep{cambcite}, for the \emph{Planck} $\Lambda$CDM cosmology \citep{planck}. We include the corrections from \cite{2015MNRAS.454.1958M} for the non-linear corrections in the matter power spectrum. In these calculations, the comoving distance at a given redshift is determined using the \texttt{astropy} package \citep{astropy1, astropy2}.

\subsection{Linear decoupled approximation test}
Here we use the simulations to test the linear decoupled approximation of equation (\ref{lindec}) compared to the full expression in equation (\ref{matrices}). In these simulated tests we use a maximum multipole of $L=64$ when calculating this full expressions, this is limited by the complexity of computing the $P$ and $Q$ terms in equation (\ref{matrices}) that scale like $L^6$ and since we are testing the linear decoupled approximation we cannot use the numerical advantages described in \cite{bct}. We also compare difference between the measured change in EE power spectrum computed using a forward model and the analytic predictions using the linear decoupled approximation alone in which case we can use a higher maximum multipole of $L=2048$.

In Figure \ref{fig:power} we show the difference between the measured EE power spectrum (computed using a forward model) and the analytic predictions using the full calculation and the linear decoupled approximation. When forward modelling we create measured shear data using equation (\ref{gamma}) and then compute measured power spectra via equations (\ref{shtransform}) and (\ref{eqsize}). We find that in all cases the difference between the analytic expressions and the forward model is at least three to four orders of magnitude smaller than the cosmic variance error, given by \citep{2008cosm.book.....W}
\begin{equation}
\label{cv}
{\mathcal V}=\left[\frac{2}{f_{\rm sky}\Delta\ell(2\ell+1)}\right]^{1/2},    
\end{equation} 
where $\Delta \ell$ is any bandwidth in $\ell$-modes used, and $f_{\rm sky}$ is the fraction of the sky observed. We also find that the difference between the full expression (equation \ref{full}), and using the linear decoupled approximation (equation \ref{lindec}), is negligible over the tested range compared to cosmic variance terms and for most modes the predictions are indistinguishable (the very small difference is attributable to numerical rounding errors caused in the sums in equation \ref{matrices}). 

Our comparison with the cosmic variance is made on a mode-by-mode basis, however in the propagation into cosmological parameter estimation a sum over all modes is performed. The residual shown between the forward model and the linear decoupled case will propagate into cosmological parameter as a residual power spectrum change of the form $2m C_{\ell}-(\delta C^A_{\ell}-\delta C^F_{\ell})C_{\ell}$ i.e. a correction term. Since previous work has found that $m\simeq 10^{-2}$--$10^{-3}$ \cite[e.g.][]{AR}, using linear decoupled assumptions, our result that $(\delta C^A_{\ell}-\delta C^F_{\ell})\approx 10^{-5}C_{\ell}$ means that any difference with the full case will lead to a negligible overall bias on cosmological parameter estimation when summed over all modes.

\subsection{Multiplicative bias tests}
\label{Multiplicative bias tests}
As described in Section \ref{pcombosec} one can either use the EE only power, or combine the likelihood of the EE and BB power to gain additional information on the multiplicative bias.

Here we test and compare these approaches on Gaussian random field simulations, we use an all-sky survey with a maximum $\ell$-mode of $L=2048$ and use $20$ logarithmic spaced bins between $[2,L]$. For the shear field we use the \emph{Planck} cosmology used in the previous section, and scale the input power spectrum with an amplitude $AC^{EE}_{\ell}$ with a fiducial value of $A=1$. For the noise field we assume $\sigma_e=0.3$ and $N_{\rm gal}=148510660n_0$ with $n_0=30$ galaxies per square arcminute as a fiducial case. After creating a Gaussian random field we then include a constant multiplicative bias, which needs to be marginalised or removed from the inference. The free parameters are $(m, \sigma_e, A)$, and in all cases we assume a Gaussian likelihood. We will show results of estimating the parameters from the Gaussian random field simulations; we use {\tt emcee} \citep{2013PASP..125..306F,corner} for the parameter estimation and use $50$,$000$ samples in each test (removing the first 100 points and using 32 walkers), we assume uniform prior ranges of $-1\leq m\leq 1$, $0\leq A\leq 2$ and $0\leq \sigma\leq 1$ except where otherwise stated.
\begin{figure*}
\centering
\includegraphics[width=0.32\columnwidth]{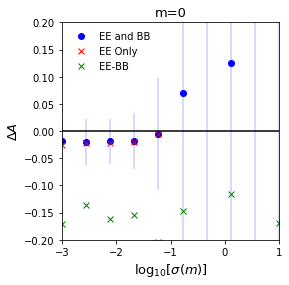}
\includegraphics[width=0.32\columnwidth]{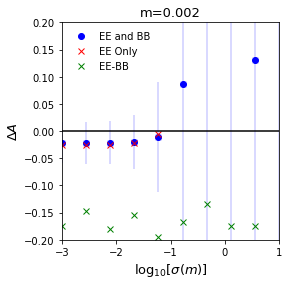}
\includegraphics[width=0.32\columnwidth]{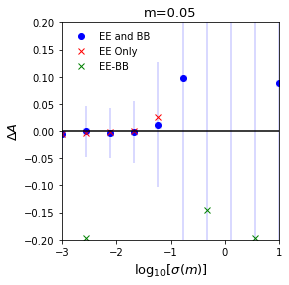}
\caption{The bias on an inferred amplitude of the amplitude parameter $A$ where the true EE power is $AC^{EE}_{\ell}$ as a function of the width of a prior distribution on the multiplicative bias $m$ (left for $m=0$, centre for $m=0.002$, and right for $m=0.05$). We use the Gaussian random field simulations described in Section \ref{Simple Simulations}. We show results for the joint EE and BB analysis (blue), EE Only (red), and EE-BB (green). The we show $1$-$\sigma$ error bars on $A$ for the joint EE and BB analysis only, these are similar for the EE-BB and EE Only points. The EE-BB is in some cases biased low by more than the plot axes, which we truncate to highlight the small remaining biases for the EE and BB, and EE Only analyses.}
\label{fig:mprior}
\end{figure*}

In Figure \ref{G2} we show the constraints when only the flat priors are used on $m$ and $\sigma_e$. In this case we find that the all three parameters are completely degenerate and no meaningful constraint on the amplitude is possible. Therefore a constraint on the cosmic shear amplitude is only possible with either a prior on $m$, a prior on $\sigma_e$ or both. If the prior on $m$ or $\sigma_e$ is too large, or centred on the incorrect value, then the constraints on $A$ will be biased and the error bar larger. 

The fact that constraints on $\sigma_e$ and $A$ are completely degenerate depends on the statistic used. For EE only we have $(1 + 2m)(A C^{EE}_{\ell} + \sigma^2_e N_l)$, and so the observed EE power only constrains the total combination of these amplitudes; $(1+2m)$ is degenerate with $A$ from the signal contribution and $\sigma^2_e$ from the noise part and hence the three parameter space is degenerate. For the EE and BB case the terms $(1+2m)\sigma^2_e$ and $(1 + 2m)(A C^{EE}_{\ell} + \sigma^2_e N_{\ell})$ are independently constrained by the EE and BB power respectively in the joint likelihood. 

In Appendix A we show that in general the marginalisation over $m$ will result in biases on the inferred amplitude of $A$, caused by a classical marginalisation paradox, but that the total amplitude of the power spectrum $(1+2\delta m)A$ should be unbiased, where $\delta m$ is a residual bias that is consistent with zero. In Figure \ref{fig:mprior} we demonstrate this by applying priors to $m$ for the three statistics we investigate (we only apply the uniform prior on $\sigma_e$ and $A$ in this case). We test this for the cases that the true value of $m=0$, $m=0.002$ and  $m=0.05$, and vary $\sigma(m)$ with the prior centred on the true value. This leads to asymptotic estimates of $A$ for small $\sigma(m)$, however these are biased low due to the marginal distribution of $A$ being biased, and this bias is larger for smaller $m$ values as shown in Appendix A. We find that the EE-BB is affected more than the EE only and joint EE and BB likelihood, because in this case there is no additional information from the stochastic ellipticity term.

To avoid the biases in the marginal distribution of $A$ we can instead characterise the total amplitude using $(1+2\delta m)A$ where $\delta m$ is a residual bias. This estimator should be an unbiased for $A$ if $\delta m$ is zero. To estimate the residual bias from two-point statistics one can do this in two ways
\begin{itemize}
    \item Measure this from simulations, such that $\delta m\rightarrow 0$ with some uncertainty $\sigma(m)$, 
    \item Infer $\delta m$ from the BB power spectrum with a sufficiently good prior on $\sigma_e$. In this case it is better to have no prior on $m$ (which may bias any inference of $(1+2\delta m)A$), and we only need to characterise a prior on $\sigma_e$, 
\end{itemize}
or going beyond two-point statistics to include higher-order or point-estimate terms may also help to lift the degeneracies. 
We therefore construct an estimator for $A$ that is 
\begin{eqnarray}
\tilde A&\simeq& (1+2\delta m)A\nonumber\\
\tilde A&\simeq& (1+2[m-\langle m\rangle])A
\end{eqnarray}
where $\langle m\rangle$ is the mean $m$ measured either from simulations or inferred from the BB power. This should be unbiased by the effect of marginalisation if there is a good estimate of $\langle m\rangle$. 

\begin{figure*}
\centering
\includegraphics[width=0.32\columnwidth]{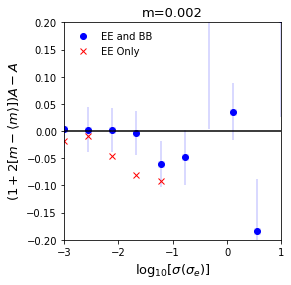}
\includegraphics[width=0.32\columnwidth]{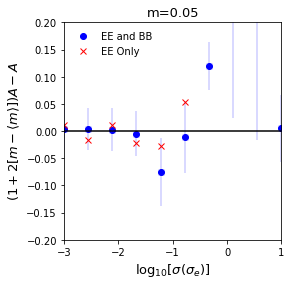}
\includegraphics[width=0.32\columnwidth]{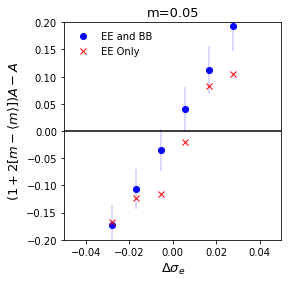}
\caption{The bias on an inferred amplitude $A$ compared to the estimator $(1+2[m-\langle m\rangle])A$ as a function of the width of a prior distribution on the stochastic ellipticity variance bias $\sigma(\sigma_e)$ (left for $m=0.002$, and centre for $m=0.05$) and $\Delta\sigma_e$ (right for $m=0.05$). We use the Gaussian random field simulations described in Section \ref{Simple Simulations}. We show results for the joint EE and BB analysis (blue) and EE Only (red). The fainter error bars are the $1$-$\sigma$ error on $(1+2m)A$ for the joint EE and BB analysis.}
\label{12m}
\end{figure*}

In Figure \ref{12m} we show the bias on the amplitude $A$ as a function of the width and centre of the prior on $\sigma_e$ and find that indeed the bias is consistent with zero when $\sigma(\sigma_e)\leq 0.05$ and $\Delta\sigma_e\leq 0.01$. We also find that this is not dependent on the overall amplitude of the bias. Constraining  $\sigma_e$ to this level may be possible using deep field observations \citep[see e.g.][]{2014MNRAS.439.1909V}. A prior on $\sigma_e$ leads to a lifting of the degeneracy in the contribution $(1+2m)\sigma^2_e$ coming from the $N_{\ell}$ contribution to the various statistics, leading to a constraint on $(1+2m)$ in the $(m,\sigma_e)$ sub-space, this in turn constrains the combination $(1+2m)A$ and hence $A$.

\begin{figure*}
\centering
\includegraphics[width=0.32\columnwidth]{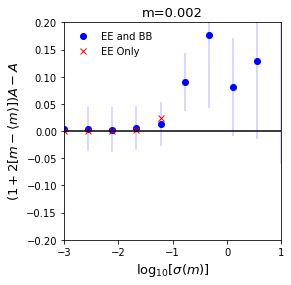}
\includegraphics[width=0.32\columnwidth]{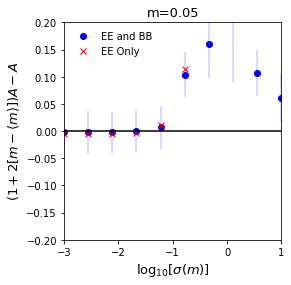}
\includegraphics[width=0.32\columnwidth]{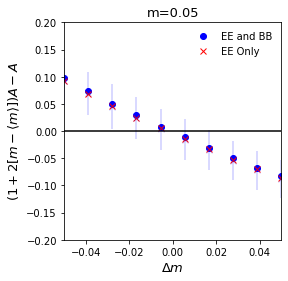}
\caption{The bias on an inferred amplitude $A$ compared to the estimator $(1+2[m-\langle m\rangle])A$ as a function of the width of a prior distribution on the multiplicative bias $\sigma(m)$ (left for $m=0.002$, and centre for $m=0.05$) and $\Delta m$ (right for $m=0.05$). We use the Gaussian random field simulations described in Section \ref{Simple Simulations}. We show results for the joint EE and BB analysis (blue) and EE Only (red). The fainter error bars are the $1$-$\sigma$ error on $(1+2m)A$ for the joint EE and BB analysis.}
\label{monly}
\end{figure*}

In Figure \ref{monly} we show the bias on the amplitude $A$ as a function of the width and centre of the prior on $m$ and also find that the bias is consistent with zero when $\sigma(m)\leq 0.07$ and $\Delta m\leq 0.01$. We also find that this is not dependent on the overall amplitude of the bias. In practice one will have a prior from calibration simulations $\Pi(m | \langle m \rangle, \sigma(m))$, and an unbiased estimate of $A$ would then be $(1+2(m-\langle m\rangle))A$ when marginalised over $m$ with the prior.

\section{Conclusions}
\label{Conclusions}
In this paper we have extended previous work to write down an expression for the propagation of multiplicative and additive weak lensing biases into cosmic shear power spectra. This expression includes terms that couple the multiplicative bias field and the survey mask, which in principle cause scale-dependent behaviour that is linear in multiplicative bias. By testing on simulations, which include some extreme cases of multiplicative bias fields, we find that the two assumptions of using only linear terms in multiplicative bias, and assuming no coupling between the bias field and the mask, are sufficient to capture any impact of multiplicative biases on cosmic shear power spectra for low-$\ell$ modes. 

In deriving this result we find several combinations of power spectra that are dependent on biases to varying degrees, and we identify that the BB power is sensitive to the multiplicative bias via the stochastic ellipticity field. We find that without prior information on either the multiplicative bias or the variance of the stochastic ellipticity that measurement of the amplitude $A$ of the cosmic shear power spectrum is completely degenerate. When applying priors to the multiplicative bias we find that this biases any inference of the amplitude parameters. However we find that the combination of $(1+2\delta m)A$ is unbiased for a joint EE and BB likelihood if the stochastic ellipticity variance is known to better than $\sigma(\sigma_e)\leq 0.05$ and $\Delta\sigma_e\leq 0.01$ or the multiplicative bias is known better than $\sigma(m)\leq 0.07$ and $\Delta m\leq 0.01$. This will be generalised to a tomographic analysis and the assessment of the bias on cosmological parameters in future work. 

\acknowledgements
{\small \emph{Acknowledgements:} TDK acknowledges funding from the European Union’s Horizon 2020 research and innovation programme under grant agreement No 776247. ACD acknowledges funding from the Royal Society. PLT acknowledges support for this work from a NASA Postdoctoral Program Fellowship. Part of the research was carried out at the Jet Propulsion Laboratory, California Institute of Technology, under a contract with the National Aeronautics and Space Administration. We thank Mark Cropper, Alan Heavens, Henk Hoekstra, Peter Schneider, and Raul Jimenez for insightful discussions. We thank the developers of {\tt SSHT}, {\tt massmappy}, and {\tt CAMB} for making their code publicly available.}

\bibliographystyle{mnras}
\bibliography{sample.bib}

\appendix

\section{Marginalisation over degenerate parameters}
\label{A:Intro}
It is well known that marginal distributions may be biased in a prior-dependent manner \citep{dawid1973marginalization} (the so-called marginalisation paradox). 

Given some angular power spectrum $C_{\ell}$ we introduce two amplitude parameters $A$ and $B$ such that 
\begin{eqnarray}
\label{clz}
\widetilde C_{\ell}=zC_{\ell}=ABC_{\ell}
\end{eqnarray}
where $\widetilde C_{\ell}$ is the modified power spectrum, $\ell$ is an angular multipole, and $z=AB$. In general if $A$ or $B$ are fixed then assuming a linear dependence on the amplitude parameters the inferred distribution of the other parameter ($B$ or $A$ respectively) will be Gaussian \citep{2019PhRvD.100b3519T}.  

However, when jointly estimating $A$ and $B$ any decrease in $A$ will be compensated in the fit by an increased $B$, and vice versa, with singularities at both $A=0$ (with $B\rightarrow\infty$) and $B=0$ (with $A\rightarrow\infty$) in this simple case. This means that in the joint fit the marginalised distributions of $A$ and $B$ are no longer Gaussian, but bounded by $[0,\infty)$. This could be generalised further where there is a constraint on the hyperbolic angle (i.e. the length of the `banana' rather than the width) as well as $z$, but in this case we consider the hyperbolic angle to be unconstrained.
\begin{figure*}
\centering
{\bf Inverse z-Gaussian Distribution}\\
\includegraphics[width=0.33\columnwidth]{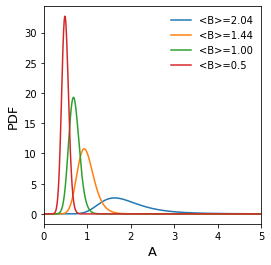}
\includegraphics[width=0.33\columnwidth]{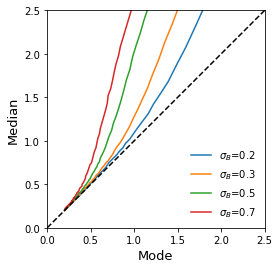}
\includegraphics[width=0.33\columnwidth]{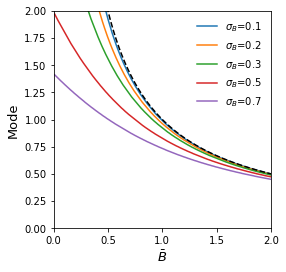}\\
{\bf z-Gaussian Distribution}\\
\includegraphics[width=0.33\columnwidth]{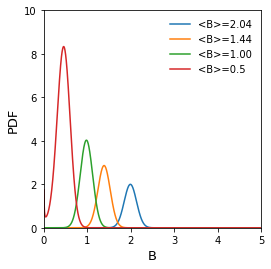}
\includegraphics[width=0.33\columnwidth]{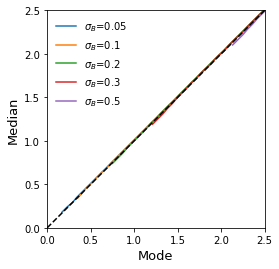}
\includegraphics[width=0.33\columnwidth]{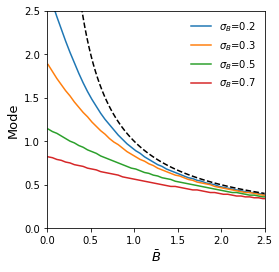}
\caption{Left: Some examples of the inverse z-Gaussian distribution (equation \ref{invzdist}, top) and z-Gaussian distribution (equation \ref{zdist}, bottom), with various mean values of $B$ with $\bar z=1$, $\sigma_z=0.1$ and $\sigma_{x}=0.1$. Centre: The median of the inverse z-Gaussian distribution compared to the mode for various values of $\sigma_B$ with $\bar z=1$, $\sigma_z=0.1$ (top) and for the z-Gaussian distribution (bottom); the black dashed lines show the case that the median is equal to the mode (i.e. the Gaussian case). Right: The mode of the distributions as a function of the centre of the Gaussian prior $\bar B$, for various $\sigma_B$. In the inverse case the dashed line is $1/\bar B$ and in the non-inverse case proportional to $\bar B$.}
\label{fig:rice1}
\end{figure*}

Assuming a Gaussian distribution in $z$ the likelihood of $z$, which is also the joint probability of $A$ and $B$ in this setup, can be written like
\begin{eqnarray}
\label{likez}
p(z)=p(A, B)=\frac{1}{(2\pi)^{1/2}\sigma_z}{\rm exp}\left(-\frac{(\bar z - AB)^2}{2\sigma^2_z}\right)
\end{eqnarray}
where $\bar z$ is the mean of $z$ and $\sigma_z$ an uncertainty. If there is prior information on one variable $\Pi(B)$, then we can construct a posterior which in the case of a Gaussian prior is  
\begin{eqnarray}
\label{post}
p(A, B)\Pi(B)=\frac{1}{(2\pi)\sigma_z\sigma_B}{\rm exp}\left(-\frac{(\bar z - AB)^2}{2\sigma^2_z}-\frac{(\bar B - B)^2}{2\sigma^2_B}\right),
\end{eqnarray}
where $\bar B$ is the mean of the prior and $\sigma_B$ the uncertainty.
The marginalised distribution of $A$ and $B$ are then given by 
\begin{eqnarray}
p(A)=\int_0^{\infty}p(A,B)\Pi(B){\rm d}B,\,\,\,{\rm and}\,\,\,p(B)=\int_0^{\infty}p(A,B)\Pi(B){\rm d}A.
\end{eqnarray}
We note that we use the most general normalisations such the probability distributions $p(z)$ and $\Pi(B)$ are normalised over $(-\infty,\infty)$, but that we only consider the marginalised distributions over $(0,\infty)$ in the case we consider because $\bar{z}>0$ and $A>0$ i.e. the overall amplitude is positive.  

In the Gaussian case the marginalised distribution is given by
\begin{eqnarray}
\label{zdist}
p(B)=\frac{1}{B(8\pi)^{1/2}\sigma_B}\left(1+{\rm erf}\left[\frac{\bar z}{\sqrt{2}\sigma_z}\right]\right)
{\rm exp}\left(-\frac{(B-\bar B)^2}{2\sigma^2_B}\right)
\end{eqnarray}
where $\bar z$ is the mean of $AB$ with some error $\sigma_z$, and we have imposed a prior on $B$ with mean $\bar B$ and error $\sigma_B$. In this case the marginalised distribution of the unconstrained parameter $A$ is 
\begin{eqnarray}
\label{invzdist}
\label{probm}
p(A)=\frac{1}{(8\pi \alpha(A,\sigma_z,\sigma_B))^{1/2}}
\left(1+{\rm erf}\left[\frac{\sigma^2_BA\bar z+\sigma^2_z\bar B}{(2\alpha(A,\sigma_z,\sigma_B))^{1/2}\sigma_z\sigma_B}\right]\right)
{\rm exp}\left(-\frac{(\bar z-\bar B A)^2}{2\alpha(A,\sigma_z,\sigma_B)}\right) 
\end{eqnarray}
where $\alpha(A,\sigma_z,\sigma_B)=A^2\sigma^2_B+\sigma^2_z$. We refer to equations (\ref{zdist}) and (\ref{invzdist}) as the z-Gaussian and inverse z-Gaussian distributions respectively (although they are not technically related via an inverse relation). Both of these are described by four free parameters ($\bar B$, $\sigma_B$, $\bar z$, $\sigma_z$).

In Figure \ref{fig:rice1} we show some examples of the z-Gaussian and inverse z-Gaussian distributions for ($\bar z=1$, $\sigma_{B}=0.1$, $\sigma_z=0.1$). We also show in Figure \ref{fig:rice1} the difference between the median and the mode/maximum of the distributions, for various values of $\sigma_{B}$ (keeping $\bar z=1$ and $\sigma_z=0.1$), where in general the mean and medians are skewed to larger values than the mode, which is much more pronounced for the inverse z-Gaussian distribution. This shows that $B$ is close to a Gaussian distribution but that $A$ is non-Gaussian. We also show the mode of the distributions compared to the centre of the Gaussian prior $\bar B$ and find that both the z-Gaussian distribution and inverse z-Gaussian modes are always biased low i.e. the z-Gaussian distribution is close to Gaussian but with a mode shifted away from the input Gaussian case. 

Therefore we conclude that if one performs parameter estimation on $z$ directly with no prior on any parameter this this should be unbiased. However when jointly fitting the degenerate free parameters $A$ and $B$ to data $A$ and $B$ will be biased towards lower values, and the mode of $z$ can be biased. 
\begin{figure*}
\centering
\includegraphics[width=0.35\columnwidth]{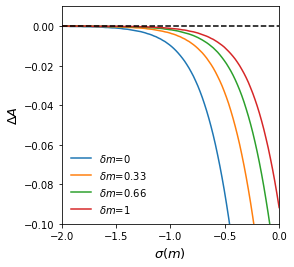}
\includegraphics[width=0.34\columnwidth]{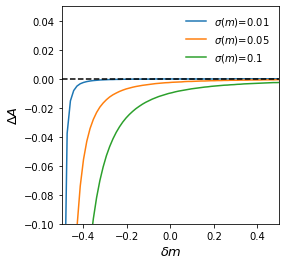}
\caption{Left: The expected change in the mode of amplitude of the cosmic shear power spectrum $\Delta A$ a function of the width of a prior on $m$, for various value of the true $m$ value assuming the prior is centred on the true value, with a fixed $\sigma_z=0.1$. Right: The expected mode of the amplitude of the cosmic shear power spectrum a function of the true $m$ for various $\sigma(m)$ values, assuming that $\sigma_z=0.1$.}
\label{fig:priorm}
\end{figure*}

\subsection{Application to Cosmic Shear}
To explore this formalism we consider the cosmic shear power spectrum (equation \ref{eq:cltheory}) where the overall amplitude of the cosmic shear power spectrum is $z\propto (1+2\delta m)A$, where we include the affect of a residual multiplicative bias $\delta m$ (see equations \ref{lindec}).

So in this case we have that $A$ is the fiducial (unbiased power spectrum) amplitude and $B=(1+2\delta m)$ will be marginalised over. For an unbiased case $\delta m=0$ and we have the $\bar B=1$, and for a biased case $\delta m>0$ we have that $\bar B>1$. Therefore we expect from the discussion in Section \ref{A:Intro} that the amplitude of the power spectrum should be biased low when such marginalisation is performed, and the bias should decrease as the true value of nuisance parameter $m$ increases.

In Figure \ref{fig:priorm} we show how the mode of amplitude will change as a function of the width of the prior on $\delta m$ for various cases of $\sigma(m)$ and the true value of $\delta m$. We find that for reasonable value of $\sigma_m$ and $\sigma_z$, for a best case that $\delta m=0$ that the mode of the marginalised amplitude value can be biased low by up to $\sim 1$--$2\%$. In all cases we assume a best-case that the prior is centred on the true value of $\delta m$, which in reality may not be the case. However the use of simulations \citep{Hoekstra17} and/or additional information from the noise (B-mode) power spectrum will enable calibration of the mean of $m$, which we explore in Section \ref{Simple Simulations}.

\label{lastpage}
\end{document}